\documentclass{article}
\usepackage{spconf,amsmath,graphicx}
\usepackage{comment}
\usepackage{algorithm}
\usepackage{algorithmic}
\usepackage{array}
\usepackage{graphicx}
\usepackage{cite}
\usepackage{multirow}
\usepackage{flushend}
\usepackage{url}

\DeclareGraphicsExtensions{.pdf,.png,.jpg}

\def\Z{{\textbf{Z}}}
\def\X{{\textbf{X}}}
\def\H{{\textbf{H}}}
\def\Zhat{{\hat{\textbf{Z}}}}
\def\Real{{\mbox{real}}}
\def\Imag{{\mbox{imag}}}

\title{A Complex Matrix Factorization approach to joint modeling of magnitude and phase for source separation}
\name{Chaitanya Ahuja, Karan Nathwani and Rajesh M. Hegde}
\address{Indian Institute of Technology, Kanpur\\
   Email: \{chahuja, nathwani, rhegde\}@iitk.ac.in}
\begin{document}
\ninept

\maketitle

\begin{abstract}
Conventional NMF methods for source separation factorize the
matrix of spectral
magnitudes. Spectral Phase is not included in the decomposition process of
these methods. However, phase of the speech mixture is generally used in
reconstructing the target speech signal. This results in undesired traces
of interfering sources in the target signal. In this paper the spectral
phase is incorporated in the decomposition process itself. Additionally,
the complex matrix factorization problem is reduced to an NMF problem using
simple transformations.  This results in effective separation of speech
mixtures since both magnitude and phase are utilized jointly in the
separation process. Improvement in source separation results  are
demonstrated using objective quality evaluations on the GRID corpus.
\end{abstract}
\section{Introduction}
\label{sec:intro}
 Monaural speaker separation is challenging in the presence
 of a competing speaker, due to all the information mixed up in a single channel. This results in degradation of intelligibility of the target speaker speech in the presence of an interfering speaker. There have been significant breakthroughs to tackle this problem in the yesteryear. Though, when compared to humans' innate ability to separate mixed speech intuitively, the separation algorithms have a long way to go. This serves as a motivation to develop such source separation systems, which can achieve performance comparable to humans.

 In literature, many source separation algorithms have been developed. Computational auditory scene analysis (CASA) \cite{bregman1994auditory}, hidden Markov models (HMM) \cite{roweis2000one}, sinusoidal modeling \cite{mowlaee2010improved} and non-negative matrix
factorization (NMF) \cite{mowlaee2010improved}.
 NMF \cite{lee1999learning,lee2001algorithms} has been
 widely used for source separation. In NMF, power spectrograms have been analyzed to reveal underlying latent components of audio signals. Other methods include modifying conventional NMF by applying sparseness constraints and achieving temporal continuity of sources\cite{virtanen2007monaural}.

\begin{figure}[!htbp]
 \centering
\includegraphics[width=8cm]{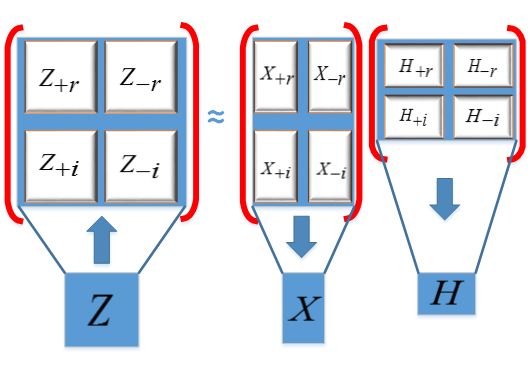} 
 \caption{Illustrating CMF for joint modeling of phase and magnitude}
 \label{fig:overview}
\end{figure}

A novel method to factorize complex matrices is proposed in this paper. This method converts the complex matrix factorization problem to a non-negative matrix factorization (NMF) problem by using simple transformations. Conventional NMF factorizes the magnitude of the input complex matrix, hence disregarding phase. Additionally, phase of the mixed signal is generally used for individual signal reconstruction which brings undesired traces of interfering sources in the target signal. In the proposed method, phase is taken into account while decomposition itself and thus is called complex matrix factorization (CMF). CMF has been attempted before in \cite{parry2007incorporating,kameoka2009complex,king2010single,kirchhoff2014towards}. Some of these methods assume a probabilistic approach while estimating the error where as our method involves a deterministic approach to solve the problem at hand.

NMF has been used for various applications other than source separation. A denoising method using NMF has been explained in \cite{wilson2008regularized}. In \cite{bertin2010enforcing} NMF has been applied to polyphonic music transcription. Speech Enhancement has also been performed using an NMF framework in  \cite{mohammadiha2013supervised}. Multi-channel source separation using factorization of complex data has been discussed in \cite{sawada2013multichannel}. We will, instead, look into application of the proposed CMF in supervised single-channel separation domain. Our proposed method converts the complex matrix to a non-negative matrix while maintaining the integrity of the problem. Hence, for all methods based on an NMF framework, CMF could be a desired alternative.

Objective evaluations on separated individual speech signals are used for illustrating the significance of the proposed method when compared to other single channel source separation methods in literature. GRID corpus database has been used in the performance evaluation.

Terminologies used throughout the paper are as follows. $|\textbf{A}|$ and $\phi_\textbf{A}$ gives the magnitude and phase respectively of a complex matrix \textbf{A}, $\|.\|$ represents the Frobenius norm in all cases.

The remainder of this paper is organized as follows. Section 2 describes
problem formulation for source separation in anechoic environment. In
Section 3, Matrix Factorization is explained along with the Complex Matrix Factorization (CMF) formulation. An algorithm is also proposed to incorporate the new theory into application. Section 4 deals with Performance Evaluation of phase reconstruction and speech separation. Finally, in Section 5 the discussion is concluded with future prospects of the proposed theory.

\section{Problem Formulation}
\label{sec:format}
Let us consider a mixed speech signal $z(n)$ consisting of two speakers
$z_1(n)$ and $z_2(n)$. The objective of speaker separation is to
obtain the estimates of $z_1(n)$ and $z_2(n)$ where $n$ are the time samples. Speech signals have huge amount of variation in time-domain, hence signals are transformed to frequency-domain for further analysis.  Let $Z(k,m), Z_1(k,m)$ and $Z_1(k,m)$
represent the STFT of $z(n)$, $z_1(n)$ and $z_2(n)$ respectively.
Here, $k$ represents frequency bin index and $m$ corresponds to the frame
index in STFT. Since STFT is linear, we can write \\
\begin{equation}
Z(k,\omega)=Z_1(k,\omega)+Z_2(k,\omega)
\end{equation}
\begin{equation}
 |Z(k,\omega)| e^{j\phi_{Z(k,\omega)}}= |Z_1(k,\omega)|
e^{j\phi_{Z_1(k,\omega)}} +  |Z_2(k,\omega)| e^{j\phi_{Z_{2}(k,\omega)}}
\end{equation}

\begin{figure*}[!htb]
\hspace{0.5cm}
  \includegraphics[width=17cm]{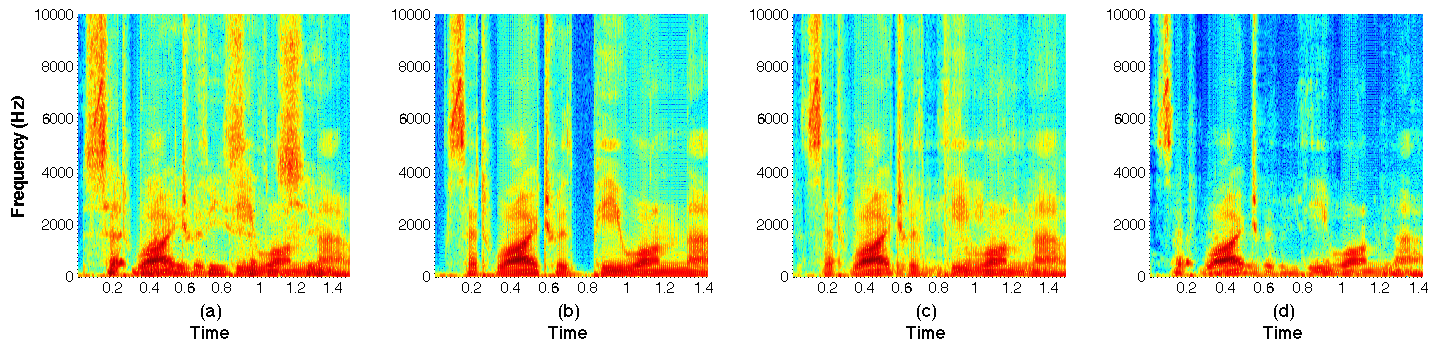}
  \caption{Spectrogram of (a) Mixed speech signal (b) Individual speech signal used as a ground truth, (c) Reconstructed speech signal via CMFbrian, (d) Reconstructed speech signal via proposed CMF }
	\label{fig:spectrogram}
\end{figure*}

Standard separation methods involve constructing trained bases \cite{6887314}  for both the speakers in question. With the constructed bases, corresponding weights are calculated for a mixture giving way to estimation of separated speech signals.

We use speech $z_i(n)$ of the $i^{th}$ speaker from the training set of clean speech to generate a bases vector set $\X_{\mbox{train}}$. This bases vector set can be used to estimate weights $\H_i$ corresponding to each speaker. Both, generating a bases vector set and estimation of weights require CMF. Hence the problem reduces to finding an accurate technique to estimate complex bases $\X_{\mbox{train}}$ and corresponding weights $\H_i$ such that $\Z_i \approx \X_{\mbox{train}}\H_i$.

\section{Complex matrix factorization approach to joint modeling of magnitude and phase}
\label{sec:factorization}
Non negative matrix factorization is a widely accepted method for single-channel source separation. Decomposition of the speech into basis vectors and corresponding weights has been shown to work well for signal-channel mixtures. In general, Non-Negative Matrix Factorization (NMF) has been used to factorize the magnitudes in the given matrix. Phase, is either taken to be equal to the input signal or is reconstructed via various methods. 

Given a Non-Negative Matrix $\Z$, we factorize it to non-negative factors $\X$ and $\H$ such that
\begin{equation}
\Z \approx \X\H
\end{equation}

This problem does not have a closed-form solution. Classically, numerical solutions have been computed by constructing an appropriate optimization problem. We have fast converging iterative algorithms which ensure reduction in distance between $\Z$ and $\X\H$ after successive updates. The proposed Complex Matrix Factorization has been formulated for Euclidean Distance metric, hence Euclidean Distance is minimized in the classic NMF domain

\begin{equation}
\label{eq:orig} 
\begin{array}{c}
\min \|\Z - \X\H\|^2 \mbox{ with respect to }\X \mbox{ and } \H\\
\\
\Z\mbox{, }\X \mbox{ and } \H \mbox{ are Non-Negative Matrices}\\
\end{array}
\end{equation}

Iterative Updates in \cite{lee2001algorithms}, that ensure convergence of $\X$ and $\H$, are given as follows
\begin{equation}
\label{eq:nmf}
  \begin {array}{cc}
    \X_{mn} \leftarrow \X_{mn} \frac{\left( \Z\H^T \right)_{mn}}{\left( \X\H\H^T \right)_{mn}} &
    \H_{np} \leftarrow \H_{np} \frac{\left( \X^T\Z \right)_{np}}{\left( \X^T\X\H \right)_{np}}
  \end{array}
\end{equation}

It has been proved in literature that every update will decrease the distance between $\Z$ and $\X\H$. Stability of the updates have also been discussed in \cite{badeau2011stability}.
 
In Section \ref{ssec:cmf}, we start with a new method of Complex Matrix Factorization (CMF)  which is used to reconstruct phase and magnitude jointly, within the NMF framework. Discussion related to the need of phase reconstruction is covered in Section \ref{sssec:phase}. Reconstruction of individual speech signals is talked about in \ref{ssec:reconstruction}. In Section \ref{ssec:algo}, an algorithm has been proposed which incorporates all the modifications.

\subsection{The proposed complex matrix factorization approach}
\label{ssec:cmf}
Consider $\Z$ to be a complex matrix. Let the bases vectors be denoted by a matrix $\X$ and the corresponding weights by $\H$. Here $\X$ is complex and $\H$ real. Also, let $\Zhat = \X\H$. To reduce CMF to NMF, we perform separation in $\Z$, $\Zhat$, $\X$ and $\H$ (also shown in \textbf{Figure \ref{fig:overview}}) via a simple transformation given as follows

\begin{equation}
  \Zhat = \Zhat_{+r}- \Zhat_{-r}+ j \left( \Zhat_{+i}-\Zhat_{-i} \right)
\end{equation}

\noindent where,

\begin{equation}
\label{eq:separation}
  \begin{array}{cc}
    \Zhat_{+r}=\max \left( 0,\Real\left( \Zhat \right) \right) & \Zhat_{-r}=-\min \left( 0,\Real\left( \Zhat \right) \right)\\
\\
    \Zhat_{+i}=\max \left( 0,\Imag\left( \Zhat \right) \right) & \Zhat_{-i}=-\min \left( 0,\Imag\left( \Zhat \right) \right)\\
\\
  \end{array}
\end{equation}

\noindent where $\max$, $\min$, $\Real$ and $\Imag$ are element-wise functions, taking maxima, taking minima, real part and imaginary part of each element. 

$\Z$ is also separated as described in Equation \ref{eq:separation}, whereas $\X$ and $\H$ are to be separated as follows

\begin{equation}
\X=\X_{+r}-\X_{-r}+j \left(\X_{+i}-\X_{-i} \right)
\end{equation}

\begin{equation}
  \H=\H_{+}-\H_{-}
\end{equation}

\noindent where $\X_{+r}$, $\X_{-r}$, $\X_{+i}$, $\X_{-i}$, $\H_{+}$ and $\H_{-}$ are non-negative matrices

Simplifying and comparing LHS and RHS of $\Zhat=\X\H$ we get 

\begin{equation}
\label{eq:comparision}
  \begin{array}{cc}
    \Zhat_1=\Zhat_{+r}=\X_{+r}\H_{+}+\X_{-r}\H_{-} \\  \Zhat_2=\Zhat_{-r}=\X_{+r}\H_{-}+\X_{-r}\H_{+}\\
    \Zhat_3=\Zhat_{+i}=\X_{+i}\H_{+}+\X_{-i}\H_{-} \\  \Zhat_4=\Zhat_{-i}=\X_{+i}\H_{-}+\X_{-i}\H_{+}\\
  \end{array}
\end{equation}

Lastly, for convenience sake let 
\begin{equation}
\label{eq:simplicity}
  \begin{array}{cccc}
    \Z_{1}=\Z_{+r} &  \Z_{2}=\Z_{-r} &  \Z_{3}=\Z_{+i} &  \Z_{4}=\Z_{-i}\\
    \
  \end{array}
\end{equation}

With all the equations in place, let us move to the transformation of CMF to NMF. Apply triangle inequality to Equation \ref{eq:orig} to get
\begin{equation}
\min_{\X ,\H} \| \Z - \X\H \|^2 \leq \min_{\X, \H} \sum_{k=1}^{4}{\| \Z_k-\Zhat_k\|^2}
\end{equation}

\noindent As $\Z_k$'s and $\Zhat_k$'s are independent of each other we get
\begin{equation}
\label{eq:solution}
\min_{\X, \H} \| \Z - \X\H \|^2 \leq  \sum_{k=1}^{4} {\min_{\X, \H} \| \Z_k-\Zhat_k\|^2}
\end{equation}

The problem now reduces to $\min_{\X, \H} \| \Z_k-\Zhat_k\|^2$ for all $k \in \lbrace 1,2,3,4\rbrace$. RHS value of Equation \ref{eq:solution} gives an upper bound to the solution of the optimization problem in Equation \ref{eq:orig}. Hence convergence of RHS of Equation \ref{eq:solution} guarantees convergence of the cost function in Equation \ref{eq:orig}. 

Now, we have 4 optimization problems to be solved simultaneously with same variables having dependencies on different cost functions. Solving them sequentially would lead to a bias towards the first optimization problem. To avoid divergent solutions, we combine the sub-matrices to get a single matrix. This is shown as follows

\begin{equation}
\label{eq:final}
  \left( \begin{array}{cc}
      \Zhat_{+r} & \Zhat_{-r} \\
      \Zhat_{+i} & \Zhat_{-i} \\
    \end{array}\right)=
  \left( \begin{array}{cc}
      \X_{+r} & \X_{-r} \\
      \X_{+i} & \X_{-i} \\
    \end{array}\right)
  \left( \begin{array}{cc}
      \H_+ & \H_- \\
      \H_- & \H_+ \\
    \end{array}\right)
\end{equation}

or,

\begin{equation}
\label{eq:final1}
  \left( \begin{array}{cc}
      \Zhat_1 & \Zhat_2 \\
      \Zhat_3 & \Zhat_4 \\
    \end{array}\right)=
  \left( \begin{array}{cc}
      \X_{1} & \X_{2} \\
      \X_{3} & \X_{4} \\
    \end{array}\right)
  \left( \begin{array}{cc}
      \H_1 & \H_2 \\
      \H_3 & \H_4 \\
    \end{array}\right)
\end{equation}

\begin{equation}
\label{eq:final2}
\Zhat_{c}=\X_{c}\H_{c}
\end{equation}

As $\H_1=\H_4$ and $\H_2=\H_3$, we perform an update after every NMF iteration which takes care of the aforementioned constraints.

\begin{equation}
\label{eq:hupdate}
\begin{array}{cc}
 \H_1,\H_4 \leftarrow \frac{\H_1+\H_4}{2} &  \H_2,\H_3 \leftarrow \frac{\H_2+\H_3}{2}\\
\end{array}
\end{equation}

The CMF problem is now reduced to an NMF problem of the form

\begin{equation}
\label{eq:problem} 
\begin{array}{c}
\min \|\Z_c - \X_c\H_c\|^2 \mbox{ with respect to }\X_c \mbox{ and } \H_c\\
\\
\Z_c=
\left(
\begin{array}{cc} \Z_{+r} & \Z_{-r}\\ 
                  \Z_{+i} & \Z_{-i}\\
\end{array}\right)
\mbox{, }\X_c \mbox{ and } \H_c \mbox{ are Non-Negative Matrices}\\
\end{array}
\end{equation}

This can be solved by various methods in literature, of which one of them is referred to in Equation \ref{eq:nmf}.

\subsubsection{Significance of phase spectrum in reconstruction of individual signals}
\label{sssec:phase}
In general, phase of the individual source signals is not used in estimating the separated signals. The original phase of the mixture is taken as it is for the reconstructed separated signal in the conventional methods \cite{smaragdis2007convolutive}. However, phase plays an important role in the reconstruction of individual source signals. This can be noted in \cite{gerkmann2012phase}, where the  estimated signal's SNR increases by up to 1.8 dB. In this work, phase is taken into account in the decomposition process itself. This leads to a robust speech reconstruction method with improved perceptual quality.

\subsection{Reconstruction of individual speech signals}
\label{ssec:reconstruction}
For the $i^{th}$ speaker, trained bases $ \X_{\mbox{train}(i)}$ are obtained by applying CMF on 

\begin{equation}
  \Z_i \approx \X_{\mbox{train}(i)}\hat{\H}_i
\end{equation}

\noindent Given a mixed speech signal $\Z$ of speaker $i$ and $j$ in STFT domain, and $ \X_{\mbox{train}(i)}$'s as known and fixed quantities, we solve for $\H_i$ and $\H_j$ by applying CMF on

\begin{equation}
  \Z \approx 
\left( \begin{array}{cc}
  \X_{\mbox{train}(i)} & \X_{\mbox{train}(j)}\\
  \end{array} \right)
\left(  \begin{array}{c}
  \H_{(i)}\\
  \H_{(j)}\\
\end{array} \right)
\end{equation}

\noindent Separated speech signals $\Z_i^{\mbox{estm}}$ and $\Z_j^{\mbox{estm}}$ are estimated by
\begin{equation}
\begin{array}{cc}
\Z_i^{\mbox{estm}} \leftarrow \X_{\mbox{train}(i)}\H_{(i)} & \Z_j^{\mbox{estm}} \leftarrow \X_{\mbox{train}(j)}\H_{(j)}\\
\end{array}
\end{equation}

\subsection{Algorithm to compute bases and weights using  the proposed CMF method}
\label{ssec:algo}
The algorithmic steps to compute the bases $\X$ and corresponding weights $\H$ are listed in Algorithm 1.
\begin{algorithm}[H]
\caption{: Algorithm to compute $\H$, $\X$ using proposed CMF Method}
\begin{algorithmic}[1]{\label{alg1}}
\STATE {\bf{Initialization}}: Random non-negative values are assigned to $\X_{+r}$, $\X_{-r}$, $\X_{+i}$, $\X_{-i}$, $\H_{+}$ and $\H_{-}$.  \\
\STATE Rearrange these sub-matrices to form $\X_c$ and $\H_c$ as shown in Equation \ref{eq:final1} and \ref{eq:final2}.\\
\STATE $\X_{c(ij)} \leftarrow \X_{c(ij)} \frac{\left( \Z_c\H_C^T \right)_{(ij)}}{\left( \X_c\H_c\H_c^T \right)_{(ij)}}$\\
\STATE $\H_{c(jk)} \leftarrow \H_{c(jk)} \frac{\left( \X_c^T\Z_c \right)_{(jk)}}{\left( \X_c^T\X_c\H_c \right)_{(jk)}}$\\
\STATE $\H_1,\H_4 \leftarrow \frac{\H_1+\H_4}{2}$ and  $\H_3,\H_2 \leftarrow \frac{\H_2+\H_3}{2}$.\\
\STATE {\bf{Repeat}}: Step 2 through 5 for a number of iterations to minimize the distance between $\Z$ and $\Z_c$.\\
\STATE {\bf{Termination}}: $\X \leftarrow \X_1-\X_2 + j \left( \X_3 -\X_4 \right)$ and $\H \leftarrow \H_+-\H_-$ to reconstruct the actual factors along with the correct phases.
\end{algorithmic}
\end{algorithm}

\begin{table*}[!htb] 
\caption{Objective Evaluation results of individual speech reconstructed by various methods } 
\centering 
\begin{tabular}{c |rr| rr| rr| rr} 
\hline\hline 

&\multicolumn{2}{c|}{NTF} & \multicolumn{2}{c|}{NMF} &\multicolumn{2}{c|}{CMFbrian} & \multicolumn{2}{c}{CMF}\\
\hline
\textbf{Methods} & $\mu$ & $\sigma$ & $\mu$ & $\sigma$ & $\mu$ & $\sigma$ & $\mu$ & $\sigma$\\
 \hline
PESQ              &    0.81  &        0.56  &    2.03  &        0.50  &      2.31  &        0.55  &    2.26  &        0.35  \\
TIRloss           &    0.96  &        0.02  &    0.96  &        0.01  &      0.89  &        0.05  &    0.89  &        0.01  \\
 TIRLESC           &    0.74  &        0.14  &    0.50  &         0.1  &      0.12  &        0.05  &  0.40  &        0.08  \\
\hline 
\end{tabular} 
\label{table:evaluation} 
\end{table*}

\section {Performance Evaluation}
\label{sec:simulations}

Section \ref{ssec:database} describes the database used for performance evaluation of the algorithm. Spectrographic Analysis and Phase reconstruction are discussed in Section \ref{ssec:estimation} and \ref{ssec:phaserec} respectively.

\subsection{Database}
\label{ssec:database}
Grid-Corpus Database \cite{cooke2006audio} is used for testing purposes in this work. This database consists of 1000 clean speech signals for each of the 34 speakers listed. Audio-Intelligibility tests indicated that speech material is understandable without the video, hence the database is used to test and compare various algorithms.

Mixtures of speech signals are generated with target to interference ratio equal to 1. The experiments are performed in a supervised manner. We use 200 speech signals of first 10 speakers for training and use 100 speech signals of the same speakers for testing. The proposed algorithm is compared with other methods in literature using the testing set.

\subsection{Spectrographic Analysis}
\label{ssec:estimation}
Training data from Grid-Corpus \cite{cooke2006audio} was used to estimate bases vectors for each speaker. The proposed algorithm in Section \ref{ssec:algo} was applied to estimate the separated signals from a given mixture of speech signals which are a part of the Testing data. A sample of reconstructed Spectrograms by the proposed CMF and CMF in \cite{king2010single} are depicted in Figure \ref{fig:spectrogram}. Demo of source separation can be seen at\footnote{\url{http://home.iitk.ac.in/~rhegde/chdemo.html}}.

\subsection{Phase Reconstruction Accuracy }
\label{ssec:phaserec}
\begin{figure}[htbp]
 \centering
\includegraphics[width=8.5cm]{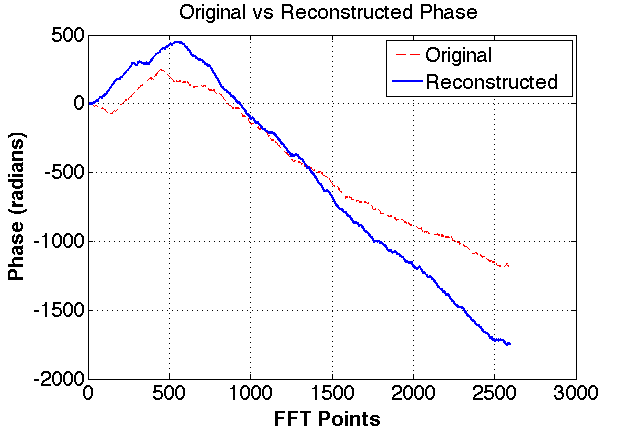} 
 \caption{Comparison of original versus estimated phase of one time-frame of STFT. The phase was estimated by the proposed CMF method}
 \label{fig:phase}
\end{figure}

Simulations were performed by factorizing STFT of some speech signals. This was done to test the convergence of Algorithm-\ref{alg1} for complex signals. Figure \ref{fig:phase} gives a pictorial representation of phase of a column vector of STFT of input ($\Z$) versus estimated phase of the respective column vector of STFT of output ($\Zhat$). 

\subsection{Objective evaluation of reconstructed speech signals}
Reconstruction was performed for 500 mixtures generated from Grid-Cropus. Non-negative matrix factorization (NMF), Non-negative tensor factorization (NTF) \cite{barker2013non}, Complex-matrix factorization in \cite{king2010single} (CMFbrian) and the proposed Complex-matrix factorization have been used on the same testing data to extract individual speech signals from a given mixture. Objective evaluation values PESQ, target to interference ratio loss (TIRLoss) and  excitation spectra correlation (TIRESC) have been calculated for all factorization methods and are listed in Table \ref{table:evaluation}. TIRLoss and TIRLESC are values similar to SNRLoss and SNRLESC defined in  \cite{Ma:2011:SLN:1937187.1937333} with the signal being replaced by the target-speaker and noise by interference.

PESQ \cite{rix2001perceptual} gives a overall speech quality evaluation on a scale of 1 (bad) to 5 (good). TIRLoss gives a quantitative value to loss due to interference on a scale of 0 (good) to (bad). TIRESC $\left(=\left[\mbox{TIRloss}\right]\left[1-r^2\right]\right)$ is also a value between 0 (good) to 1 (bad), where $r$ is the correlation coefficient between the clean speech and reconstructed speech of the target speaker.

The mean scores ($\mu$) obtained, imply that CMF performs much better than NTF and NMF. It performs equally well when compared to CMFbrian. The standard deviation ($\sigma$) of PESQ and TIRloss values of reconstructed speech by CMF is lower than CMFbrian which indicates that the performance of CMF remains more consistent than CMFbrian. Although, the reconstructions by CMF and CMFbrian are competitive, the proposed CMF is computationally more efficient as it uses the standard NMF framework.

\section{Conclusion}
\label{sec:conclusion}

A new method of complex matrix factorization, which jointly utilizes
both the spectral magnitude and phase is proposed in this work for single
channel source separation. In this work the phase spectrum is incorporated
into the decomposition stage, along with magnitude, making it a complex
factorization method. Additional contributions of this work include
converting the complex matrix factorization method into a standard NMF
method using simple transformations.

Its superiority is demonstrated with respect to other methods, using
magnitude only reconstruction, motivating the need for incorporating phase
into the decomposition process.  Although this method has been
applied to single-channel source separation, the proposed algorithm
and can be applied to any generalized NMF method with applications in
speech enhancement, music transcription and multi channel source separation. Currently we are investigating different distance measures to obtain better performance at lower SNR.

\vfill\pagebreak

\bibliographystyle{IEEE}
\bibliography{strings,ref}

\end{document}